\documentclass[a4paper,11pt]{article}
\pdfoutput=1 

\usepackage{jcappub} 

\usepackage[T1]{fontenc} 
\usepackage{epstopdf}  
\usepackage{ulem}   

\definecolor{grey}{rgb}{0.1,0.1,0.6}

\title{Deflection angle and lensing signature of covariant $f(T)$ gravity
}

\author[a,b,c]{Xin Ren,}
\author[a,b,c]{Yaqi Zhao,}
\author[d,a,b,1]{Emmanuel N. Saridakis,}
\author[a,b,c,1]{Yi-Fu Cai\note{Corresponding author}}

\affiliation[a]{Department of Astronomy, School of Physical Sciences,\\
University of Science and Technology of China, 96 Jinzhai Road, Hefei, Anhui 230026, China}
\affiliation[b]{CAS Key Laboratory for Research in Galaxies and Cosmology,\\
University of Science and Technology of China, 96 Jinzhai Road, Hefei, Anhui 230026, China}
\affiliation[c]{School of Astronomy and Space Science, \\
University of Science and Technology of China, 96 Jinzhai Road, Hefei, Anhui 230026, China}
\affiliation[d]{National Observatory of Athens, Lofos Nymfon, 11852 Athens, Greece}

\emailAdd{rx76@mail.ustc.edu.cn}
\emailAdd{zxmyg86400@mail.ustc.edu.cn}
\emailAdd{msaridak@noa.gr}
\emailAdd{yifucai@ustc.edu.cn}

\abstract{
We calculate the deflection angle, as well as the positions and magnifications of the lensed images, in the case of covariant $f(T)$ gravity. We first extract the spherically symmetric solutions for both the pure-tetrad and the covariant formulation of the theory, since considering spherical solutions the extension to the latter is crucial,  in order for the results not to suffer from frame-dependent artifacts. 
Applying the weak-field, perturbative approximation we extract the deviations of the solutions comparing to General Relativity. Furthermore, we calculate the deflection angle and then the differences of the positions and magnifications in the lensing framework. This effect of consistent $f(T)$ gravity on the lensing features can serve as an observable signature in the realistic cases where $f(T)$ is expected to deviate only slightly from General Relativity, since lensing scales in general are not restricted as in the case of Solar System data, and therefore deviations from General Relativity could be observed more easily.
}

\begin{document}

\maketitle

\section{Introduction}
 
Modified gravity offers an alternative explanation to the two phases of accelerated expansion of the universe \cite{Capozziello:2011et, Nojiri:2010wj}, and it can also improve the renormalizability of General Relativity and thus may bring gravity closer to a quantum description \cite{Stelle:1976gc}. The usual approach towards the construction of   modified theories of gravity is to add various extra terms in the Einstein-Hilbert action   \cite{Clifton:2011jh, DeFelice:2010aj}. However, one can alternatively start from the equivalent formulation of gravity in terms of torsion, namely the Teleparallel Equivalent of General Relativity (TEGR) \cite{Unzicker:2005in, Aldrovandi:2013wha, Maluf:2013gaa}, and extend the corresponding Lagrangian, namely the torsion scalar $T$, in various ways, resulting in $f(T)$ gravity \cite{Cai:2015emx, Bengochea:2008gz, Linder:2010py}, in $f(T,T_G)$ gravity \cite{Kofinas:2014owa, Kofinas:2014daa}, $f(T,B)$ gravity \cite{Bahamonde:2015zma, Karpathopoulos:2017arc, Boehmer:2021aji}, scalar-torsion theories \cite{Geng:2011aj, Hohmann:2018rwf, Bahamonde:2019shr}, etc. Torsional gravity, and in particular $f(T)$ theory, has been shown to lead to interesting behavior, both in cosmological applications \cite{Chen:2010va, Zheng:2010am, Bamba:2010wb, Cai:2011tc, Capozziello:2011hj, Wei:2011aa, Amoros:2013nxa, Otalora:2013dsa, Bamba:2013jqa, Li:2013xea, Paliathanasis:2014iva, Malekjani:2016mtm, Farrugia:2016qqe, Qi:2017xzl, Cai:2018rzd, Abedi:2018lkr, El-Zant:2018bsc, Anagnostopoulos:2019miu, Nunes:2019bjq, Cai:2019bdh, Yan:2019gbw, ElHanafy:2019zhr, Saridakis:2019qwt, Wang:2020zfv, Hashim:2020sez, Ren:2021tfi}, as well as in the case of spherically symmetric solutions \cite{Boehmer:2011gw, Gonzalez:2011dr, Ferraro:2011ks, Wang:2011xf, Atazadeh:2012am, Rodrigues:2013ifa, Nashed:2013bfa, Nashed:2014iua, Junior:2015fya, Kofinas:2015hla, Das:2015gwa, Rani:2016gnl, Rodrigues:2016uor, Mai:2017riq, Singh:2019ykp, Nashed:2020kjh, Bhatti:2018fsc, Ashraf:2020yyo, Ditta:2021wfl}.

Spherically symmetric solutions are in general of great importance for the experimental tests of a gravity theory \cite{Will:1993ns}. According to Birkhoff's theorem, in General Relativity  the exterior solution of a spherical, non-rotating, gravitating body must be given by the Schwarzschild metric \cite{Meng:2011ne}, and this lies as a basis for the Solar System test of General Relativity. Concerning torsional gravity, spherically symmetric solutions have been used in order to extract Solar System constraints on $f(T)$ gravity \cite{Iorio:2012cm, Xie:2013vua, Iorio:2016sqy} and on $f(T,B)$ gravity \cite{Bahamonde:2020bbc, Farrugia:2020fcu}.

In the usual formulation of $f(T)$ gravity one sets the spin connection to zero, applying the so-called pure-tetrad formalism. However, such a simplification, although efficient in the case of Cartesian Friedmann-Robertson-Walker metric in cosmological applications, it may lead to undesirable features in the case of spherically symmetric solution, since only specific tetrads should be used, namely the ones that are consistent with the zero spin connection. Nevertheless, if one uses the covariant formulation of $f(T)$ gravity, namely the one that uses both the tetrad and the spin connection in order to have theoretical consistency \cite{Krssak:2015oua, Krssak:2018ywd, Hohmann:2019nat}, then the above ambiguities disappear, since every tetrad choice is equally good.

In the present work we are interested in investigating the spherically symmetric solutions and the corresponding observational consequences on the light deflection angle, but applying the consistent, covariant version of $f(T)$ gravity. In this way we can ensure that the results do not suffer from frame-dependent artifacts that appear in the pure-tetrad formalism. We first extract the corrections to the spherical solutions  compared to General Relativity in the weak-field limit, and then we calculate the deviations in the induced light deflection angle. Then we apply the result to the lensing framework \cite{Bartelmann:1999yn, Hoekstra:2008db, Bartelmann:2010fz, Cunha:2018acu,Younas:2015sva}, and we find the effect of $f(T)$ gravity on the positions and magnifications of the lensed images. As we show, this could serve as the basis for realistically observing a signature of covariant $f(T)$ gravity, since lensing scales in general are not  restricted as the Solar System ones, and therefore deviations from General Relativity could be observed more easily.

The outline of this paper is as follows. In Section \ref{sec:theory review}, we briefly review covariant $f(T)$ gravity. In Section \ref{sec:equation and solution}, we extract spherically symmetric solutions in the weak-field limit, in both the pure-tetrad and in the covariant formulations of the theory. In Section \ref{sec:deflection angle} we expand the tetrad solutions and  calculate the bending angle, and then we use this expression to find the differences in the positions and magnifications of the lensed images in comparison to General Relativity. Finally, Section \ref{sec:conclusion} is devoted to discussion and conclusions.

\section{$f(T)$ gravity}
\label{sec:theory review}

In this Section we briefly review $f(T)$ gravity, both in its simple, as well 
as in its covariant formulation. In torsional formulation the dynamical 
variable is the tetrad field $h^{a}{}_{\mu}$, which is an orthonormal basis for the tangent space at each point $x^\mu$ of the manifold.   Greek indices correspond to the spacetime coordinates and Latin indices correspond to the tangent space coordinates. Additionally, the relation between tetrad and the manifold metric is
\begin{equation}
	g_{\mu \nu}=\eta_{a b} h^{a}{}_{\mu} h^{b}{}_{\nu} ~,
\end{equation}
where $\eta_{A B}=(1,-1,-1,-1)$.
Concerning the connection, in teleparallel geometry one introduces the 
Weitzenb\"{o}ck one, $\hat{\Gamma}^{\lambda}{}_{\mu \nu} \equiv h_{a}{}^{\lambda} \partial_{\nu} h^{a}{}_{\mu}$, where for the moment we consider a vanishing spin connection \cite{Aldrovandi:2013wha,Maluf:2013gaa}. Hence, the torsion tensor is given as
\begin{equation}
    \label{eq:T_0}
   \hat{T}^{\lambda}{}_{\mu \nu} \equiv \hat{\Gamma}^{\lambda}{}_{\nu \mu}-\hat{\Gamma}^{\lambda}{}_{\mu \nu}=h_{a}{}^{\lambda}\left(\partial_{\mu} h^{a}{}_{\nu}-\partial_{\nu} h^{a}{}_{\mu}\right) ~. 
\end{equation}
One can proceed to the construction of the torsion scalar $\hat{T}$ by considering 
suitable contractions of the above torsion tensor. Nevertheless, one should 
note that in this way the theory  loses Lorentz invariance \cite{Cai:2015emx}. 
Hence, a more consistent, covariant formulation would be needed, instead of the above  "pure tetrad teleparallel gravity" \cite{Krssak:2018ywd}, namely one that is based on the complete torsion tensor \cite{Krssak:2015oua}
\begin{equation}
    \label{eq:T_spin}
    {T}^{\lambda}{}_{\mu \nu}=h_{a}{}^{\lambda}(\partial_{\mu} h^{a}{}_{ 
\nu}-\partial_{\nu} h^{a}{}_{\mu}+{\omega}^{a}{}_{b \mu} 
h^{b}{}_{\nu}-{\omega}^{a}{}_{b \nu} h^{b}{}_{ \mu}) ~,
\end{equation}
where the ${\omega}^{a}{}_{b \mu}$ is spin connection which represents 
inertial effects. 

Defining the contortion tensor as 
\begin{equation}
	K^{\rho}{}_{\mu \nu} \equiv \frac{1}{2} \Big( T_{\mu}{}^{\rho}{}_{\nu} +{T_{\nu}{}^{\rho}}_{\mu} -{T^{\rho}}_{\mu \nu} \Big) ~,
\end{equation}
and the super-potential as 
\begin{equation}
	S_{\rho}{}^{\mu \nu} \equiv \frac{1}{2} \Big( {K}^{\mu, \nu}{}_{\rho} +\delta_{\rho}^{\mu} T^{\alpha \nu}{}_{\alpha} -\delta_{\rho}^{\nu} T^{\alpha \mu}{}_{\alpha} \Big) ~,
\end{equation}
we can now construct the torsion scalar through the contraction 
\begin{equation}
    \label{Tscalardef}
    T = S_{\rho}{}^{\mu \nu} T^{\rho }{}_{\mu \nu} ~.
\end{equation}
Thus, the action of the Teleparallel Equivalent of General Relativity (TEGR) 
is written as
\begin{equation}
	S=\int d^{4} x \frac{h}{16\pi G} \big[ T+ {\mathcal{L}}_m \big] ~,
\end{equation}
where $h=\det\left(h^{a}{}_{\mu}\right)=\sqrt{-g}$, and variation in terms of 
the tetrads gives rise to exactly the same equations with General Relativity.

In TEGR, the contributions of the spin connection in the field equation cancels each other in a specific way and the spin connection has no effect on the final field equation \cite{Aldrovandi:2013wha, Maluf:2013gaa,Krssak:2015lba,Krssak:2015rqa}. The corresponding Lagrangian can be re-written as 
\begin{equation}
\label{eq:lrs}
    {\mathcal{L}} \left( h^{a}{}_{\mu}, {\omega}^{a}{}_{b \mu} \right) = {\mathcal{L}} \left(h^{a}{}_{\mu}, 0 \right) + \frac{1}{\kappa} \partial_{\mu} \left( h {\omega}^{\mu} \right) ~,
\end{equation}
where ${\omega}^{\mu}={\omega}^{ab }{}_{\nu}h_{a}{}^{\nu}h_{b}{}^{\mu}$.
Hence, spin connection contributes only to the boundary term. In TEGR, the ${\mathcal{L}} \left( h^{a}{}_{\mu}, {\omega}^{a}{}_{b \mu} \right)$ is complete torsion scalar $T$ \eqref{Tscalardef}, while the ${\mathcal{L}} \left(h^{a}{}_{\mu}, 0 \right)$ is the non-complete torsion scalar $\hat{T}$ obtained by the torsion tensor \eqref{eq:T_0} with vanishing spin connection
\begin{align}
\label{Tspure}
    \hat{T} = \hat{S}_{\rho}{}^{\mu \nu} \hat{T}^{\rho }{}_{\mu \nu} .
\end{align}
Nevertheless, when one extends the TEGR theory to $f(T)$ gravity, the situation becomes different. In particular, the function of the total divergence term will not be a total divergence term anymore. Similarly to the fact that $f(T)$ gravity is not equivalent to $f(R)$ gravity, the torsion tensor with or without the spin connection leads to a different theory. In other words, the spin connection ${\omega}^{\mu}$ which ensures the covariance of the theory will appear in the field equations \cite{Krssak:2015oua, Hohmann:2018rwf, BeltranJimenez:2018vdo, Golovnev:2020las}.

We use the switching off gravity method to determine the spin connection \cite{Krssak:2015oua}. The spin connection can be expressed as 
\begin{equation}
\label{eq:spin}
    {\omega}_{\ b \mu}^{a}=\frac{1}{2} h_{(\mathrm{r})\mu}^{c} \left[f_{b}{}^{a}{}_c \left(h_{(\mathrm{r})}\right)+f_{c}{}^{a}{}_b \left(h_{(\mathrm{r})}\right)-f_{\ b c}^{a} \left(h_{(\mathrm{r})}\right)\right] ~,
\end{equation}
where $h_{(r)\mu}^a$ is the reference tetrad with gravity switched off, which can be obtained by setting the gravitational constant $G$ equal to zero, i.e. $h_{(r)\mu}^a=h^{a}{}_{\mu}\mid_{G=0}$ \cite{Krssak:2018ywd}. Moreover, note that we have introduced the coefficients of anholonomy $f^{c}{}_{a b}$, which represent the curls (infinitesimal circulations) of the basis members, and can be expressed as \cite{Aldrovandi:2013wha, Maluf:2013gaa}
\begin{equation}
    f^{c}{}_{a b}=h_{a}{}^{\mu} h_{b}{}^{ \nu}\left(\partial_{\nu} 
h^{c}{}_{\mu}-\partial_{\mu} h^{c}{}_{\nu}\right) ~.
\end{equation}
Accordingly, using a non-zero spin connection one obtains the fully covariant version of the theory. We mention here that the switching-off method is not fundamental, the determination of spin connection is discussed further in \cite{Golovnev:2017dox,Emtsova:2021ehh,Golovnev:2021lki}.

\section{Spherically symmetric solutions}
\label{sec:equation and solution}

In this Section we proceed to the extraction of spherically symmetric solutions 
in $f(T)$ gravity. In particular, we desire to obtain the spherically symmetric metric 
\begin{equation}
\label{metricdef}
 d s^{2}=A(r)^2 d t^{2}-B(r)^2 d r^{2}-r^{2} d \Omega^{2} ~.
\end{equation}
As we mentioned, although in the case of TEGR all tetrad choices that correspond to the above metric are valid choices, in the case of pure-tetrad $f(T)$ gravity one should suitably choose the tetrad ansatz in order to maintain theoretical consistency. On the contrary, this is not the case in the covariant version of the theory, in which all tetrad choices are valid as long as they are accompanied by the suitable spin connection choice. For completeness, in the following we will examine the two cases separately.

\subsection{Pure-tetrad $f(T)$ gravity}
\label{sec:ptg}

The action of pure-tetrad $f(T)$ gravity is
\begin{equation}
 S = \int d^{4} x \frac{h}{16\pi G} [f(\hat{T}) + {\mathcal{L}}_m] ~.
\end{equation}
It is worth noting that, under pure-tetrad theory, $\hat{T}$ in the action is a non-complete torsion scalar. Since the spin connection has been imposed to vanish, the field equations are derived through variation  with respect to  the tetrad, leading to
\begin{equation}
\label{eq:feq}
 f_{T} \partial_{\nu}\left(h h_{a}{}^{\rho} \hat{S}_{\rho}{}^{\mu \nu}\right) +h\left(f_{T T}  h_{a}{}^{\rho} \hat{S}_{\rho}{}^{\mu \nu} \partial_{\nu} \hat{T}-f_{T}h_{a}{}^{\rho} \hat{T}^{\lambda}{}_{\nu \rho} \hat{S}_{\lambda}{}^{\nu \mu} +\frac{1}{4} f h_{a}{}^{\mu}\right) = 4\pi G h h_{a}{}^{\rho} \overset{(m)}{T}_{\rho}{}^{\mu} ~,
\end{equation}
where the hat symbols denoting quantities calculated in pure-tetrad formalism, 
and with the $T$-subscripts denoting derivatives with respect to $T$. Additionally, $\overset{(m)}{T}_{\rho}{}^{\mu}$ stands for the energy-momentum tensor of the matter sector.

As we mentioned above, in pure-tetrad $f(T)$ gravity, in the case of spherical symmetry, the diagonal tetrad choice is not a consistent one, unless $f_{TT}=0$, in which case we recover TEGR. The general tetrad could be derived through symmetry principles as \cite{Tamanini:2012hg, Hohmann:2019nat, Krssak:2018ywd}:
\begin{equation}
\label{eq:gte}
\footnotesize{h^{a}{ }_{\mu} =
 \left(
 \begin{array}{cccc}
 A(r)  & 0 & 0 & 0 
 \\ 
 0 & B(r)   \sin \theta \cos \phi & -r(\cos \theta \cos \phi \sin \gamma+\sin \phi \cos \gamma) & r \sin \theta(\sin \phi \sin \gamma-\cos \theta \cos \phi \cos \gamma) 
 \\
 0 & B(r)   \sin \theta \sin \phi  & r(\cos \phi \cos \gamma-\cos \theta \sin \phi \sin \gamma) &-r \sin \theta(\cos \theta \sin \phi \cos \gamma+\cos \phi \sin \gamma)
 \\
 0 & B(r)   \cos \theta & r \sin \theta \sin \gamma& r \sin ^{2} \theta \cos \gamma
 \end{array}
 \right)
 } ~.
\end{equation}
 We mention that in this case the general field equations (\ref{eq:feq}) provide also a non-trivial non-diagonal equation, namely   $f_{TT} \hat{T}' \cos\gamma=0$, which can be non-trivially satisfied (i.e. without setting $f_{TT}=0$) by imposing $\cos\gamma=0$.

\subsubsection{Specific tetrad Case A} 
\label{sub:case A}

Let us first consider the subcase where $\cos\gamma=0$ is realized alongside 
$\sin\gamma=1$. Hence, the tetrad ansatz for this Case A is 
\cite{Tamanini:2012hg, Ruggiero:2015oka}:
\begin{equation}
 h^{a}{ }_{\mu}=
 \left(
 \begin{array}{cccc}
 A(r) & 0 & 0 & 0
 \\
 0 & B(r) \sin (\theta ) \cos (\phi ) & -r \cos (\theta ) \cos (\phi ) & r \sin (\theta ) \sin (\phi )
 \\
 0 & B(r) \sin (\theta ) \sin (\phi ) & -r \cos (\theta ) \sin (\phi ) & -r \sin (\theta ) \cos (\phi )
 \\
 0 & B(r) \cos (\theta ) & r \sin (\theta ) & 0
 \\
 \end{array}
 \right) ~.
 \label{caseAchoice}
\end{equation}
The torsion scalar is easily found to be
\begin{equation}
 \label{eq:tf1}
 \hat{T}_A=\frac{2 (B+1) [ 2 r A'+A (B+1) ]}{r^2 A B^2} ~,
\end{equation}
while the nontrivial components of the field equations are
\begin{align}
\label{eq:a1}
 & A \Big[ 4 f_{T} r B'+f r^2 B^3-4 B^2 (f_{T}+f_{TT} r \hat{T}') -4 B (f_{T}+f_{TT} r \hat{T}') \Big] -4 f_{T} r B (B+1) A' = 0 ~, \\
\label{eq:a2}
 & 4 f_{T} r (B+2) A'+A \big( -f r^2 B^2 +4 f_{T} B +4 f_{T} \big) = 0 ~, \\
\label{eq:a3}
 & 2 r \Big\{-f_{T} r A' B'+B [A' (3 f_{T}+f_{TT} r \hat{T}')+f_{T} r A'' ] + 2 f_{T} B^2 A' \Big\} \nonumber \\
 & +A \Big[ -2 f_{T} r B'+B^3 ( 2 f_{T}-f r^2 ) +2 B^2 (2 f_{T}+f_{TT} r \hat{T}') + 2 B (f_{T}+f_{TT} r \hat{T}') \Big] = 0 ~.
    \end{align}

When $f(\hat{T})=\hat{T}$, the above tetrad ansatz gives the Schwarzschild solution. For a general $f(T)$ function, one cannot always extract an exact analytic solution. Nevertheless, when $r$ is sufficiently large, applying the weak-field approximation one can analytically extract the leading order of the metric solution approximately.  We follow the weak-field method in \cite{ Ruggiero:2015oka}, expanding the tetrad functions up to second order as 
\begin{align}
\label{eq:expand}
 & A(r)^2 \approx 1 -\frac{{p_1}}{r} +\frac{{p_2}}{r^2} +\mathcal{O}\Big(\frac{1}{r^3}\Big) ~, \nonumber \\
 & B(r)^2 \approx 1 +\frac{{p_1}}{r} +\frac{{p_3}}{r^2} +\mathcal{O}\Big(\frac{1}{r^3}\Big) ~,
\end{align} 
with $p_1$, $p_2$, $p_3$ constants. Then the torsion scalar is approximated as 
\begin{equation}
 \hat{T}_A=\frac{8}{r^2}-\frac{5 {p_1}^2 +16p_2+8p_3}{2 r^4} ~.
\end{equation}

It is worth mentioning that there are two ways to obtain  $B(r)$ here, $+\sqrt{B(r)^2}$ and $-\sqrt{B(r)^2}$. We choose the positive one in the following calculations, which implies that one can re-obtain the Minkowski case with $ (A \to 1, B \to 1).$ The other case will be discussed in Sec.\ref{covariant}. It is known that viable modified gravity models, and viable $f(T)$ models in 
particular, should be small deviations from General Relativity \cite{Nesseris:2013jea}. Hence, one can impose the parameterization $f(\hat{T})=\hat{T}+g(\hat{T})$ which quantifies the departure from General Relativity, where 
$g(\hat{T})$ is the small deviation. Since $\hat{T}$ is small in the case of weak gravitational fields we can expand $g(\hat{T})$ in powers of $\hat{T}$, i.e. we can write $f(\hat{T})\approx \hat{T} +\alpha \hat{T}^2+\mathcal{O}(\hat{T}^3)$ (actually this is what one should acquire from the effective field theory approach to torsional gravity \cite{Li:2018ixg, Chen:2019ftv}).
Therefore, inserting the above weak-field  approximations into \eqref{eq:a1}-\eqref{eq:a3} we obtain the weak-field solution of spherically symmetric case as:
\begin{align}
 & A(r)^2 \approx 1 -\frac{2M}{r} -\frac{32\alpha}{r^2} +\mathcal{O}\Big(\frac{1}{r^3} \Big) ~, \nonumber\\
 & B(r)^2 \approx 1 +\frac{2M}{r} +\frac{96\alpha}{r^2} +\mathcal{O}\Big(\frac{1}{r^3}\Big) ~,
\end{align}
where, for simplicity, we have set Newton's constant to $G=1$. Moreover, $2M$ arises from the integration constant. As a result, we have obtained the correction to the Schwarzschild solution arisen from the $f(T)$ modification  of Case A. This is same with the result in \cite{ Ruggiero:2015oka}.

\subsubsection{Specific tetrad Case B} 
\label{sub:case B}

Let us now investigate in the same way the subcase  where $\cos\gamma=0$ is realized alongside $\sin\gamma=-1$. The tetrad ansatz for this Case B is:
 \begin{equation}
    h^{a}{ }_{\mu}=
    \left(
    \begin{array}{cccc}
        A(r) & 0 & 0 & 0
    \\
        0 & B(r) \sin (\theta ) \cos (\phi ) & r \cos (\theta ) \cos (\phi ) & 
-r \sin (\theta ) \sin (\phi )
    \\
        0 & B(r) \sin (\theta ) \sin (\phi ) & r \cos (\theta ) \sin (\phi ) & r 
\sin (\theta ) \cos (\phi )
    \\
        0 & B(r) \cos (\theta ) & -r \sin (\theta ) & 0
    \\
    \end{array}
    \right) ~.
        \label{caseBchoice}
\end{equation}
The torsion scalar becomes
\begin{equation}
\label{eq:tfb}
 \hat{T}_B=\frac{2 (B-1) \left[A (B-1)-2 r A'\right]}{r^2 A B^2} ~,
\end{equation}
while the field equations are
\begin{align}
\label{eq:b1}
 & A \left[4 f_{T} r B' +f r^2 B^3+4 B^2 (f_{T}\!+\!f_{TT} r \hat{T}')-4 B (f_{T} + f_{TT} r \hat{T}') \right] + 4 f_{T} r (B-1) B A' =0 ~, \\
\label{eq:b2}
 & 4 f_{T} r (B-2) A'+A \left(f r^2 B^2+4 f_{T} B-4 f_{T}\right) = 0 ~, \\
\label{eq:b3}
 & 2 r \left\{ -f_{T} r A' B'+B \left[ A' (3 f_{T}+f_{TT} r \hat{T}') +f_{T} r A'' \right] -2 f_{T} B^2 A' \right\} \nonumber \\
 & - A \left[2 f_{T} r B'+B^3 \left(f r^2-2 f_{T}\right)+2 B^2 (2 f_{T} +f_{TT} r \hat{T}')-2 B (f_{T}+f_{TT} r \hat{T}') \right] =0 ~.
\end{align}
Applying the weak-field approximation \eqref{eq:expand}, the torsion scalar is approximated as $\hat{T} = \hat{T}_B +\mathcal{O}(\frac{1}{r^3})$, with
\begin{equation}
 \hat{T}_B = -\frac{{p_1}^2}{2 r^4} -\frac{{p_1}^3-4p_1p_2}{2 r^5} ~.
\end{equation}
Interestingly enough, we observe that the torsion scalar expression for 
this Case B tetrad choice has a leading term of the order $O\left(\frac{1}{r^4}\right)$, hence under the ansatz $f(\hat{T})\approx \hat{T} 
+\alpha \hat{T}^2 +\mathcal{O}(\hat{T}^3)$ equations \eqref{eq:b1}-\eqref{eq:b3} yield the solution:
\begin{align}
\label{eq:br}
 & A(r)^2 \approx 1 -\frac{2M}{r} +\mathcal{O}\Big(\frac{1}{r^3}\Big) ~, 
\nonumber\\
 & B(r)^2 \approx 1 +\frac{2M}{r} +\mathcal{O}\Big(\frac{1}{r^3}\Big) ~.
\end{align}
Therefore, Case B gives the same solution with General Relativity in the weak-field approximation, since the corrections appear only at higher orders. The fact that different tetrad choices give rise to different solutions is expected in the pure-tetrad choice and it is a disadvantage of the theory comparing with the covariant generalization.

\subsection{Covariant $f(T)$ gravity }\label{covariant}

In this subsection we proceed to the extraction of spherically symmetric solutions in the case of covariant $f(T)$ gravity, namely in the case where we 
consider a non-zero spin connection. Using the full complete torsion tensor $T$ \eqref{eq:T_spin} then the field equation becomes \cite{Krssak:2015oua}
\begin{align}
\label{eq:feq2}
 & h\left(f_{T T} h_{a}{}^{\rho} {S}_{\rho}{}^{\mu \nu} \partial_{\nu} {T} -f_{T}h_{a}{}^{\rho} {T}^{\lambda}{}_{\nu \rho} {S}_{\lambda}{}^{\nu \mu} +f_{T} \omega^{b}_{\ a \nu} h_{b}{}^{\rho} {S}_{\rho}{}^{\nu \mu} +\frac{1}{2} f h_{a}{}^{\mu} \right) \nonumber \\
 & +f_{T} \partial_{\nu}\left(h h_{a}{}^{\rho} {S}_{\rho}{}^{\mu \nu}\right) = 4\pi G h h_{a}{}^{\rho} \overset{(m)}{T}_{\rho}{}^{\mu} ~,
\end{align}
and as we observe they explicitly include the spin connection $\omega^{b}_{\ a \nu}$. Hence, one can now impose a tetrad choice at will, and use \eqref{eq:spin} in 
order to calculate the components of the spin connection.

In particular, imposing the diagonal tetrad $h^{a}{}_{\rho} = \operatorname{diag}(A(r), B(r), r, r \sin \theta)$ one finds 
\begin{equation}
 \omega^{\hat{1}}{}_{\hat{2} \theta}=-1 , \quad \omega^{\hat{1}}{}_{\hat{3} \phi} = -\sin \theta, \quad \omega^{\hat{2}}{}_{\hat{3} \phi}=-\cos \theta ~,
\end{equation}
with all other components being zero (the hat represents   indices in the tangent space). Similarly, for the Case A tetrad choice 
(\ref{caseAchoice}) the spin connection components become  
\begin{eqnarray}
   && \omega^{\hat{1}}{}_{\hat{2} \phi}=2 \sin ^2\theta , \quad 
\omega^{\hat{1}}{}_{\hat{3} \theta}=-2 \cos \phi , \quad 
\omega^{\hat{1}}{}_{\hat{3} \phi}=\sin 2 \theta  \sin \phi  ~, \nonumber
    \\
   && \omega^{\hat{2}}{}_{\hat{3} \theta}=-2 \sin \phi,  \quad 
\omega^{\hat{2}}{}_{\hat{3} \phi}=-2 \sin \theta \cos \theta \cos \phi ~.
\end{eqnarray}
Finally, for the Case B tetrad choice \eqref{caseBchoice} the components of the spin connection are all vanishing. In all cases, by substituting the tetrad ansatz alongside the corresponding spin connection into the scalar torsion expression \eqref{Tscalardef} we obtain the same result, namely
\begin{equation}
\label{eq:tf2}
 T=\hat{T}_B=\frac{2 (B-1) \left[A (B-1)-2 r A'\right]}{r^2 A B^2} ~,
\end{equation}

We mention here that the torsion scalar \eqref{eq:tf2} is the complete torsion scalar $T$ \eqref{Tscalardef}, i.e. the  one that does not change according to the choice of reference frame for the general tetrad \eqref{eq:gte} and spin connection \eqref{eq:spin}, with $ (A \to 1, B \to 1)$ in order to recover Minkowski spacetime. The complete torsion scalar $T$  corresponds to  ${\mathcal{L}} \left( h^{a}{}_{\mu}, {\omega}^{a}{}_{b \mu} \right)$ in \eqref{eq:lrs}, while the incomplete torsion scalar $\hat{T}$ \eqref{Tspure} corresponds to ${\mathcal{L}} \left( h^{a}{}_{\mu}, 0 \right)$. Although they differ by only one boundary term and give the same physics under TEGR, they give different physics in the more general $f(T)$ theory. $T \left( h^{a}{}_{\mu}, {\omega}^{a}{}_{b \mu} \right)$  is the complete form of torsion scalar, 
which is obviously expected in the covariant formulation of $f(T)$ gravity, and 
reveals its theoretical advantage. The complete torsion scalar \eqref{eq:tf2} will vanish in Minkowski spacetime $ (A \to 1, B \to 1)$. This is a property that the other incomplete forms do not exhibit. 

The weak-field method considered in Sec. \ref{sec:ptg} cannot provide  an exact Schwarzschild solution. This method can give the solution up to the first few terms of the series expansion, nevertheless 
it is difficult to acquire the higher-order terms. Eq.\eqref{eq:br} show that the differences of this $f(T)$ model solution and Schwarzschild metric appear at high order. Hence, we need a more precise way to solve the equation.

We  proceed to the perturbative solution of the above field equations. This perturbative method 
considers small deviations form TEGR, i.e. $f(T)=T+\epsilon g(T)$,  with $\epsilon $   a small tracking parameter. Hence, the solution for the tetrad functions will be a small correction to the standard Schwarzschild solution, namely \cite{DeBenedictis:2016aze, Bahamonde:2019zea, Bahamonde:2020bbc, Bahamonde:2020vpb, Pfeifer:2021njm,Pfeifer:2021njm} 
\begin{eqnarray}
\label{eq:ps1}
 &&
 A(r)^2 = 1 -\frac{2 M}{r}+\epsilon a(r) ~,
 \\
 &&B(r)^2 = \left( 1-\frac{2 M}{r} \right)^{-1} +\epsilon b(r) ~.
\end{eqnarray}
Additionally, concerning $g(T)$ we consider the quadratic correction (valid in 
the weak-field case), namely we impose $g(T) \approx \alpha T^2 +\mathcal{O}(T^3)$. Setting for convenience  
$\sqrt{\frac{1}{1-\frac{2 M}{r}}}\to x$, and keeping only  first-order 
terms in $\epsilon$, substituting all the above into equations \eqref{eq:b1}-\eqref{eq:b3}, we acquire the differential equations for $a(r)$ 
and $b(r)$ with respect to $r$ as
\begin{eqnarray}
\label{eq:c1}
 && r^2 x (2 M-r)^3 b'(r)-r x b(r) (r-2 M)^2 (2 M+r) \nonumber \\
 && -8 \alpha \left[5 M^2 (5 x-2)-6 M r (4 x-3)+6 r^2 (x-1)\right]=0 ~,
\end{eqnarray}
\begin{eqnarray}
\label{eq:c2}
 &&	 2 M r^4 a'(r)-r^5 a'(r)+2 M r^3 a(r)+r^2 b(r) (r-2 M)^2-16 \alpha r^2 \nonumber \\
 && +32 \alpha M^2 x-8 \alpha M^2-48 \alpha M r x+32 \alpha M r+16 \alpha r^2 x = 0 ~,
\end{eqnarray}
\begin{eqnarray}
\label{eq:c3}
 && 4 M r^6 x a''(r) -4 M^2 r^5 x a''(r) +r^7 (-x) a''(r) -6 M^2 r^4 x a'(r) +5 M r^5 x a'(r) \nonumber \\
 && -r^6 x a'(r)+2 M r^3 x a(r) (M-r)+8 M^4 r^2 x b'(r)-20 M^3 r^3 x b'(r) +18 M^2 r^4 x b'(r) \nonumber \\
 && -7 M r^5 x b'(r) +r^6 x b'(r) -2 M r x b(r) (M-r) (r-2 M)^2 -256 \alpha M^3 x + 80 \alpha M^3 \nonumber \\
 && +384 \alpha M^2 r x-240 \alpha M^2 r -192 \alpha M r^2 x +160 \alpha M r^2+32 \alpha r^3 x-32 \alpha r^3=0 ~,
\end{eqnarray}
two out of which are independent.
The solutions of the above equations are
\begin{eqnarray}
    \label{eq:psr}
	 &&
   \!\!\!\!\!\!\!\!\!\! \!\!\!\!\!\!\!
        a(r)=-\frac{   M^2 r^2 \left[c_2 (2 M-r)+c_1\right] }{M^2 r^3} \nonumber\\
&&\,
-\frac{2 \alpha\! \left[3 
M^3\!+\!2 M^2 r (9\!-\!32 x)+M r^2 (64 x\!+\!3)\!-\!16 r^3 x\right]\!-\!3 
\alpha r^2 (r\!-\!3 M) \ln \!
\left(\frac{r}{r-2 M}\right)}{3 M^2 r^3}
    \\ 
    &&   \!\!\!\!\!\!\!\!\!\! \!\!\!\!\!\!\!
    b(r)=\frac{  c_1 M r^3 }{  
M r^2 (r-2 M)^2}\nonumber\\
&&\,
-
\frac{ 2 \alpha \left[48 M^3 x \!+\!M^2 r (75\!-\!136 x)\!+\!M 
r^2 (88 x\!-\!69)\!-\!16 r^3 x\right]-3 \alpha r^3 \ln \!\left(\frac{r}{r-2 
M}\right)}{3 
M r^2 (r-2 M)^2}
,
\end{eqnarray}
where $c_1$ and $c_2$ are integration constants.
This solution is consistent with the result in \cite{DeBenedictis:2016aze, Bahamonde:2019zea}. 
In particular, under the variable transformation
$\mu =\sqrt{1-\frac{2 M}{r}}$, the form in the previous article can be obtained.
The choice of the coefficients and the setting of the parameters of the perturbation 
solution are not exactly the same, i.e.  there is going to be a coefficient difference. Expanding in powers of $\frac{1}{r}$, we obtain
\begin{eqnarray}
\label{eq:psr}
 && \!\!\!\!\!\!\!\!\!\!\!\!\!\!\!\!\!\!
 a(r) \approx\left(c_2+\frac{32 \alpha}{3 M^2}\right)- \left( 2 c_2 
M+c_1+\frac{32 \alpha}{M}\right)\frac{1}{r}-\frac{8}{5}  \alpha 
M^3 
\left(\frac{1}{r}\right)^5  +\mathcal{O}\Big(\frac{1}{r}\Big)^6   
\nonumber,
    \\
    &&
    \!\!\!\!\!\!\!\!\!\!\!\!\!\!\!\!\!\!
    b(r)
     \approx\frac{1}{r} \left(c_1+\frac{32 \alpha}{3 
M}\right)+ \left(4 c_1 M+\frac{128 
\alpha}{3}\right) \left(\frac{1}{r}\right)^2 
+
\left(12 c_1 M^2+128 \alpha M\right)\left(\frac{1}{r}\right)^3  \nonumber
    \\
&&
    + \left(32 c_1 M^3+\frac{1024 \alpha 
M^2}{3}\right) \left(\frac{1}{r}\right)^4    
+
\left(80 c_1 M^4+\frac{2584 \alpha 
M^3}{3}\right)\left(\frac{1}{r}\right)^5 
    +\mathcal{O}\Big(\frac{1}{r}\Big)^6 ~.
\end{eqnarray}
In order for this solution to be able to recover the vacuum limit, the integration  constants should satisfy $c_1=-\frac{32 \alpha}{3 M}$ and $c_2=-\frac{32 \alpha}{3 M^2}$. Interestingly enough, substituting these expressions into the above solutions leads to the simple result
\begin{align}
\label{eq:psr}
 a(r) & \approx -\frac{8\alpha M^3}{5r^5} +\mathcal{O}\Big(\frac{1}{r^6}\Big) ~, \nonumber \\
 b(r) & \approx \frac{8 \alpha M^3}{r^5} +\mathcal{O}\Big(\frac{1}{r^6}\Big) ~.
\end{align}

As we can see, in the weak-field limit the difference between this solution and the standard Schwarzschild one is of the order $\frac{1}{r^5}$. This implies that in order to distinguish viable $f(T)$ gravity from General Relativity in the Solar System (where the gravitational field is weak) one should have observations with very high degree of precision, which verifies the results of \cite{Iorio:2012cm, Xie:2013vua, Iorio:2016sqy}. Nevertheless, one could find a measurable effect and distinguish $f(T)$ gravity from TEGR even for small corrections in the lensing features. 

Before closing  this section  we would like to add some comments.
As we saw, what we can determine are the properties of the metric $B(r)^2$,
while $B(r)$ could be valued in more than one way.
In the above calculation  we chose $B(r)=+\sqrt{B(r)^2}$. 
However, there is another case, namely $B(r)=-\sqrt{B(r)^2}$.
These two cases actually represent different ways of returning 
to Minkowski spacetime. If we choose the negative case, 
we re-obtain Minkowski spacetime for $(A \to 1, B \to -1)$.
It is worth noting that the difference between $\hat{T}_A$ and $\hat{T}_B$ is that $B(r) \to -B(r)$. 
When we choose $(A \to 1, B \to -1)$ in order to go back to Minkowski spacetime, the
Case A would correspond to vanishing spin connection.
Definitely,  this is
only a sign difference and  does not change the physics. 
When we insert the concrete values of $A(r)$ and $B(r)$ in the $B(r)=-\sqrt{B(r)^2}$ case, 
the torsion scalar and the metric solution are the same as in the $B(r)=+\sqrt{B(r)^2}$ case.

\section{Deﬂection angle and lensing behavior}
\label{sec:deflection angle}

In this Section we provide the framework where a signature of $f(T)$ gravity 
could be seen even in the realistic cases where $f(T)$ is close to TEGR. As it is known, the propagation of light in different theories of gravity is  different, hence we start our analysis by calculating the light bending.
 
\subsection{Deﬂection angle}
  
Let us investigate the deflection of light rays propagating in a static spherically symmetric spacetime in the framework of $f(T)$ gravity. We follow  \cite{Keeton:2005jd} and extend the calculation to higher orders. Considering the metric \eqref{metricdef}, the bending angle of light $\hat{\varepsilon}$ is  
expressed as 
\begin{align}
\label{eq:general bending angle}
    \hat{\varepsilon}\left(r_{0}\right)
&
    =2 \int_{r_{0}}^{\infty}\left|\frac{d \varphi}{d r}\right| \mathrm{d} r-\pi \nonumber
    \\
&
    =2 \int_{r_{0}}^{\infty} \frac{1}{r^{2}} \sqrt{\frac{A^2 B^2}{1 / b^{2}-A^2 / r^{2}}} \mathrm{~d} r-\pi ~,
\end{align}
where   $r_0$ is the radial coordinate when the light ray is closest to the 
central mass. Furthermore, $b$ is the impact parameter, which is an invariant of the light ray, related to $r_0$ through $\frac{1}{b^{2}} =\frac{A\left(r_{0}\right)^2}{r_{0}^{2}}$ \cite{Keeton:2005jd}. It should be noted that $r_0$ and therefore $b$ are assumed small compared to the mass of the compact object $M$. Therefore, in the spacetime region we are interested in, the metric elements can be expanded as
\begin{eqnarray}
    \label{eq:PPN metric}
 && \!\!\!\!\!\!\!\!\!\!\!
    A(r)^2=1-a_{1}\left(\frac{M}{r}\right)+ a_{2}\left(\frac{M}{r}\right)^{2}- 
a_{3}\left(\frac{M}{r}\right)^{3}+ a_{4}\left(\frac{M}{r}\right)^{4}- 
a_{5}\left(\frac{M}{r}\right)^{5}\cdots
    \vspace{0.1cm}
    \\
  && \!\!\!\!\!\!\!\! \!\!\! B(r)^2=1+2 b_{1}\left(\frac{M}{r}\right)+4 
b_{2}\left(\frac{M}{r}\right)^{2}+8 b_{3}\left(\frac{M}{r}\right)^{3}+16 
b_{4}\left(\frac{M}{r}\right)^{4}+32 b_{5}\left(\frac{M}{r}\right)^{5}\cdots.
 \end{eqnarray}
Substituting $A(r)$ and $B(r)$ into eq.(\ref{eq:general bending angle}), 
expanding in powers of $\frac{M}{r}$, integrating term by term, and replacing $r_0$ in terms of $b$, we extract the bending angle as
\begin{equation}
    \label{eq:PPN bending angle}
\hat{\varepsilon}(b)   
\approx A_{1}\left(\frac{M}{b}\right)+A_{2}\left(\frac{M}{b} \right)^ 
{2}+A_{3}\left(\frac{M}{b}\right)^{3}+A_{4}\left(\frac{M}{b}\right)^{4}+A_{5} 
\left(\frac{M}{b}\right)^{5}+\mathcal{O}\left(\frac{M}{b}\right)^{6} ~,
\end{equation}
where
\begin{eqnarray}
  && 
  \!\!\!\!\!\!\!\!\!\!\!\!\!\!\!\!
  A_{1}=a_{1}+2 b_{1}, 
\nonumber\\
   &&    \!\!\!\!\!\!\!\!\!\!\!\!\!\!\!\!A_{2}=\frac{1}{4} \left(2 a_1^2+2 a_1 
b_1-2 a_2-{b_1}^2+4 {b_2}\right) \pi, 
\nonumber\\
   &&     \!\!\!\!\!\!\!\!\!\!\!\!\!\!\!\!A_{3}=\frac{35 {a_1}^3}{12}+\frac{5 
{a_1}^2 {b_1}}{2}-{a_1} \left(5 {a_2}+{b_1}^2-4 {b_2}\right)+\frac{2}{3} 
\left(3 {a_3}-3 {a_2} {b_1}+{b_1}^3-4 {b_1} {b_2}+8 {b_3}\right),
\nonumber\\
   &&     \!\!\!\!\!\!\!\!\!\!\!\!\!\!\!\!
    A_{4}=\frac{3}{64} \pi  \Big[40 {a_1}^4+32 {a_1}^3 {b_1}-12 {a_1}^2 \left(8 
{a_2}+{b_1}^2-4 {b_2}\right) \nonumber
\\
  && \ \ \ \ \ \ \, +8 {a_1} \left(-6 {a_2} {b_1}+6 {a_3}+{b_1}^3-4 {b_1} 
{b_2}+8 {b_3}\right)+24 {a_2}^2+8 {a_2} \left({b_1}^2-4 {b_2}\right) \nonumber
\\
   &&\ \ \ \ \ \ \,  +16 {a_3} {b_1}-16 {a_4}-5 {b_1}^4+24 {b_1}^2 {b_2}-32 
{b_1} {b_3}-16 {b_2}^2+64 {b_4}\Big] ~,
\nonumber\\
   &&     
   \!\!\!\!\!\!\!\!\!\!\!\!\!\!\!\!
    A_{5}
    =\frac{1001 {a_1}^5}{80}+ \frac{77 {a_1}^4 {b_1}}{8}-\frac{7}{2} 
{a_1}^3 \left(11 {a_2}+{b_1}^2-4 {b_2}\right)\nonumber
\\
     &&  
+\frac{7}{3} {a_1}^2 \left(-9 {a_2} 
{b_1}+9 {a_3}+{b_1}^3-4 {b_1} {b_2} +8 {b_3}\right)\nonumber
\\
     &&  
     +\frac{1}{3} {a_1} 
    \Big[63 {a_2}^2+14 {a_2} \left({b_1}^2-4 {b_2}\right)+28 {a_3} {b_1}-28 
{a_4}-5 {b_1}^4
\nonumber
\\
     && \ \ \ \ \ \ \ \ \,
     +24 {b_1}^2 {b_2}-32 {b_1} {b_3}-16 {b_2}^2+64 
{b_4}\Big]\nonumber
\\   &&  
    +\frac{2}{15} \Big[35 {a_2}^2 {b_1}-10 {a_2} \left(7 {a_3}+{b_1}^3-4 {b_1} 
{b_2}+8 {b_3}\right)-10 {a_3} {b_1}^2+40 {a_3} {b_2}-20 {a_4} {b_1}\nonumber
\\     && \ \ \ \ \ \  \ 
  +20 
{a_5}  +7 {b_1}^5-40 {b_1}^3 {b_2}+48 {b_1}^2 {b_3}+48 {b_1} {b_2}^2-64 {b_1} 
{b_4}-64 {b_2} {b_3}+128 {b_5}\Big] ~.
\end{eqnarray}

We mention here that the metric (\ref{eq:PPN metric}) reproduces the TEGR and 
thus General Relativity case  for
\begin{equation}
    \label{eq:grppn}
    \begin{array}{l}
        a_{1}=2 ,\quad a_{2}=0 ,\quad a_{3}=0 ,\quad a_{4}=0 ,\quad a_{5}=0,
    \\
        b_{1}=1 ,\quad b_{2}=1 ,\quad b_{3}=1 ,\quad b_{4}=1 ,\quad b_{5}=1,
    \end{array}
\end{equation}
in which case the   bending angle takes the standard form
\begin{equation}
    \label{eq:ppnangle}
    \hat{\varepsilon}_{GR}(b)
    \approx
    4 \left(\frac{M}{b}\right)+\frac{15 \pi  }{4} 
\left(\frac{M}{b}\right)^2+\frac{128 }{3}\left(\frac{M}{b}\right)^3+\frac{3465 
\pi  }{64}\left(\frac{M}{b}\right)^4+\frac{3584 
}{5}\left(\frac{M}{b}\right)^5+\mathcal{O}\left(\frac{M}{b}\right)^{6} ~.
\end{equation}

Let us now apply the general expression (\ref{eq:PPN bending angle}) for the 
solution  (\ref{eq:PPN metric})  and (\ref{eq:ps1}) that we obtained for the
covariant $f(T)$ gravity in the weak-field case  $f(T)=T+\epsilon \alpha T^2$,
i.e for the solution with 
\begin{equation}
    \begin{array}{l}
    \label{eq:ftppn}
        a_{1}=2 ,\quad a_{2}=0 ,\quad a_{3}=0 ,\quad a_{4}=0 ,\quad 
a_{5}=\frac{8 \alpha}{5 M^2}, 
    \\
        b_{1}=1 ,\quad b_{2}=1 ,\quad b_{3}=1 ,\quad b_{4}=1 ,\quad 
b_{5}=1+\frac{\alpha}{4 M^2}. 
    \end{array}
\end{equation}
We obtain
\begin{eqnarray}
   && 
   \!\!\!\!\!\!\!\!\!\!\!\!\!\!\!\!\!\!\!\!\!\!\!\!\!\!\!
   \hat{\varepsilon}_{f(T)}(b)\approx
    4 \left(\frac{M}{b}\right)+\frac{15 \pi  
}{4}\left(\frac{M}{b}\right)^2+\frac{128 
}{3}\left(\frac{M}{b}\right)^3+\frac{3465 \pi  
}{64}\left(\frac{M}{b}\right)^4
\nonumber\\
&&
+\left(\frac{3584 }{5}+\frac{128 \alpha}{15 
M^2}\right)
\left(\frac{M}{b}\right)^5+\mathcal{O}\left(\frac{M}{b}\right)^{6} ~.
\end{eqnarray}
Thus, we deduce that the difference of the deflection angle in the case of 
covariant $f(T)$ gravity and General Relativity is  
\begin{equation}   
\Delta\hat{\varepsilon}(b)=\hat{\varepsilon}_{f(T)}(b)-\hat{\varepsilon}_{GR}
(b)
\approx
\frac{128 \alpha}{15 
M^2}\left(\frac{M}{b}\right)^5+ \mathcal{O}\left(\frac{M}{b}\right)^6 ~.
\end{equation}
As mentioned above, the discrepancy appears at a very high order.
This is consistent with the calculation results of the  related recent works \cite{Bahamonde:2020bbc,Pfeifer:2021njm}, which expand deflection angle in terms of $\frac{M}{r}$. We are expanding the  deflection angle in terms of  $\frac{M}{b}$, in order  to make it easier to calculate the gravitational lensing effects. The forms of the two results can be transformed by a simple redefinition of variables.

\subsection{Lensing framework}\label{sec:lensing}

We now turn to the basic calculation of the present work, namely to find the 
effect on gravitational lensing. In particular, the change in the deflection angle affects the lensing features. Hence, knowing the result of the previous subsection, we can now examine  the position and magnification of the gravitational lensing in the case of the covariant $f(T)$ gravity. 

\begin{figure}[ht]
\centering
\includegraphics[width=3.3in]{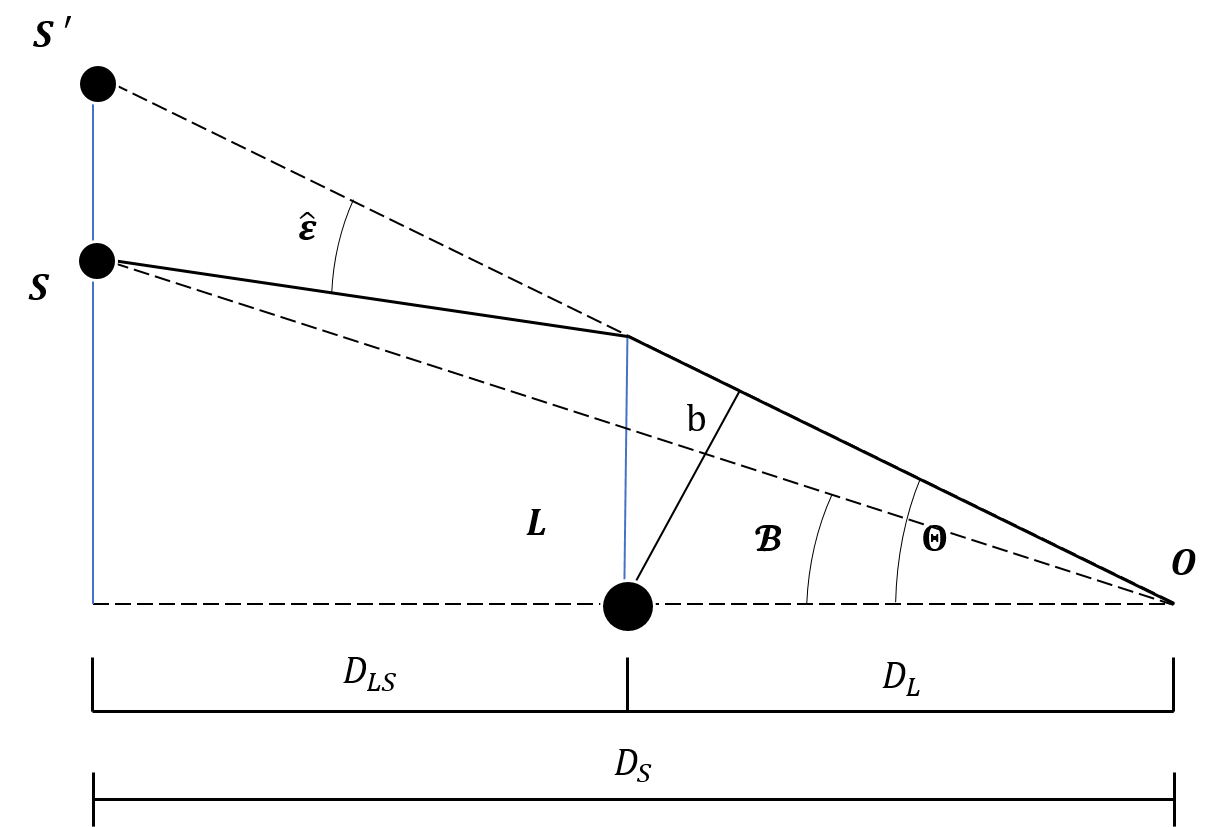}
\caption{{\it{Schematic representation of the geometric features of lensing.}}
\label{fig:lensing}}
\end{figure}

The geometrical principle of gravitational lensing is depicted in Fig.~\ref{fig:lensing}. $\mathcal{B}$ and $\Theta$ are the angular position 
of the source $S$ and the image $S'$. Moreover, $D_S$, $D_L$, $D_{LS}$ are the angular diameter distance from the observer to the light source, from the observer to the lens object, and from the lens object to the light source, respectively. Additionally, $\varepsilon$ is the bending angle and $b$ is the impact parameter discussed in the previous subsection. Finally, $\Theta$ and $b$ have the geometric relation $\Theta=\sin^{-1}\left(\frac{b}{D_L}\right)$ 
\cite{Keeton:2005jd, Ruggiero:2016iaq}.

The Virbhadra-Ellis lens equation can be expressed as \cite{Virbhadra:1999nm,Virbhadra:2002ju}
\begin{equation}
\label{eq:lens}
 \tan \mathcal{B}=\tan \Theta-D[\tan \Theta+\tan (\hat{\alpha}-\Theta)] ~,
\end{equation}
where $D=\frac{D_{LS}}{D_S}$. As usual, one introduces the Einstein angle as the standard scale for lenses, expressed as 
\begin{equation}
\theta_{E}\equiv\sqrt{\frac{4 M D_{L S}}{D_{L} D_{S}}},
\end{equation}
which allows to define the scaled parameters 
\begin{equation}
    \beta=\frac{\mathcal{B}}{\theta_{E}}, \quad \theta=\frac{\Theta}{\theta_{E}}, \quad \varepsilon=\frac{\Theta_{M}}{\theta_{E}}=\frac{\theta_{E}}{4 D},
\end{equation}
where   $\Theta_{M}=\tan^{-1}\left(\frac{M}{D_L}\right)$ is the angle relative 
to the gravitational radius of the lens. Note that
$\varepsilon$ will be 
the   expansion parameter. Finally, we fix the source position $\beta$ and 
expand the image position parameter $\theta$ as 
\begin{equation}
    \theta\approx
    \theta_{0}+\theta_{1} \varepsilon+\theta_{2} 
\varepsilon^{2}+\theta_{3} \varepsilon^{3}+\theta_{4} 
\varepsilon^{4}+\mathcal{O}(\varepsilon^{5}) ~.
\end{equation}
Using the above definition, we can now write the bending angle in the Parameterized Post 
Newtonian metric \eqref{eq:ppnangle} as
\begin{eqnarray}
&&\!\!\!\!\!\! 
    \hat{\varepsilon}=
    \frac{A_{1}  }{\theta_0}\varepsilon+\frac{ A_{2}-A_{1} 
\theta_1}{\theta_0^2}\varepsilon ^2+\frac{1}{\theta_{0}^{3}}\left[A_{3}-2 A_{2} 
\theta_{1}+A_{1}\left(\frac{8}{3} D^{2} \theta_{0}^{4} 
+\theta_{1}^{2}-\theta_{0} \theta_{2}\right)\right] \varepsilon^{3} \nonumber
    \\
&& \ \   
    +\frac{1}{3 \theta_0^4}\Big[8 A_{1}D^2 \theta_0^4\theta_1  -3 A_{1} 
\theta_0^2 \theta_3 + 6 A_{1} \theta_0\theta_1 \theta_2-3 A_{1} \theta_1^3+ 16 
A_{2} D^2\theta_0^4
\nonumber\\
&& \ \ \ \  \ \ \ \ \ \ \,  
- 6 A_{2} \theta_0\theta_2+9 A_{2}\theta_1^2-9 A_{3} 
\theta_1+3 A_{4}\Big]\varepsilon ^4 \nonumber
    \\
&&\ \   
    +\frac{1}{45 \theta_0^5} \Big\{A_1 \left\{224 D^4 \theta_0^8+120 D^2 
\theta_0^4 (\theta_0 \theta_2\!+\!2)+45\! \left[2 \theta_0^2 \theta_1 
\theta_3\!+\!\theta_0^2 \left(\theta_2^2\!-\!\theta_0 \theta_4\right)\!-\!3 
\theta_0 \theta_1^2 \theta_2\!+\!\theta_1^4\right]\right\}
\nonumber
    \\
&&\ \ \ \ \ \ \ \ \ \ \ \ 
 -45\! \left[2 A_2 \theta_0^2  \theta_3\!-\!6 A_2 \theta_0 \theta_1 
\theta_2\!+\!4 A_2 \theta_1^3\!
-\!A_3 \left(8 D^2 \theta_0^4\!-\!3 \theta_0 \theta_2\!+\!6 
\theta_1^2\right)\!+\!4 A_4 \theta_1\right]\!+\!45 A_5\Big\}\varepsilon 
^5
\nonumber
    \\
&&\ \  
+\mathcal{O}\left(\varepsilon ^6\right) ~. 
\label{eq:bend}
\end{eqnarray}

We proceed by expanding the lens equation \eqref{eq:lens} in terms of 
$\varepsilon$, and solving for the corresponding coefficients of $\theta$   order by order. The first term gives the relation between  $\beta$ and $\theta_0$, namely
\begin{align}
 \beta = \frac{4 \theta_0^2-A_1}{4 \theta_0 } ~.
\end{align}
There are two solutions of $\theta_0$, corresponding to the two images on 
either side of the source. We keep the image positions be positive. The image 
on the same side of the source $(\beta > 0)$ is labeled as $\theta_{0}^{+}$, 
while that  on different side $(\beta < 0)$ is labeled as $\theta_{0}^{-}$. Therefore, we have
\begin{align}
 \theta_{0}^{\pm}=\frac{1}{2}\left(\sqrt{A_1+\beta^{2}} \pm|\beta|\right) ~.
\end{align}
Hence, we can apply the above expression and calculate the coefficients $\theta_i$ of image  position for  the case of General Relativity, i.e. for \eqref{eq:grppn}, and for the case of covariant $f(T)$ gravity  with $f(T)=T+\alpha T^2$, i.e. for \eqref{eq:ftppn} (note that the difference in the deflection angle between the two theories occurs in $A_5$-terms). Hence, we conclude that the deviation of the image position between the two theories is   found to be 
\begin{align}
    \Delta\theta
   \approx
   \theta_{f(T)}-\theta_{GR}=\frac{32 \alpha}{15 \theta_0^3 
\left(\theta_0^2+1\right) M^2}\varepsilon^4+\mathcal{O}(\varepsilon^5) ~.
\end{align}

Finally, let us calculate the magnification of the lensed image $\mu$ at 
the position $\Theta$, which is given by \cite{Hoekstra:2008db}
\begin{align}
 \mu(\Theta)=\left[\frac{\sin B(\Theta)}{\sin \Theta} \frac{d B(\Theta)}{d \Theta}\right]^{-1} ~.
\end{align}
Since $\Theta=\theta \theta_{E}$ and $\theta$ was expanded in $\varepsilon$,
our obtained $\theta$-solution above leads to
\begin{eqnarray}
&& 
\!\!\!\!\!\!\!\!\!\!\!\!\!\!\!\!
\mu=
 \mu_{0}+\mu_{1} \varepsilon+\mu_{2} \varepsilon^{2}+\mu_{3} 
\varepsilon^{3}+\mu_{4} \varepsilon^{4}+\mathcal{O}(\varepsilon^{5}) \nonumber
  \\
&&  \!\!\!\!\!\!\!\!\!\!\!
    =\frac{16 \theta_{0}^{4}}{16 \theta_{0}^{4}-A_{1}^{2}}-\frac{16 A_{2} 
\theta_{0}^{3}}{\left(A_{1}+4 \theta_{0}^{2}\right)^{3}}\varepsilon 
\nonumber \\ 
&& \!\!\!\!
+\frac{1}{3  [ (A_1-4 \theta_0^2 )  (A_1+4 \theta_0^2 )^5 ]}\bigg\{ 8   \Big\{\theta_0^2  \big[A_1^6 D^2    
+8 A_1^5  (9 D^2-6 D-2 ) \theta_0^2-384 A_1 A_3 \theta_0^4
\nonumber \\ 
&& \ \ \  \ \ \  \ \ \  \ \ \ \ \ \  \ \ \  \ \ \  \ \,   
+32 A_1^4  (17 D^2-12 D-4 ) \theta_0^4+128 A_1^3  (9 D^2-6 D-2 ) \theta_0^6 \nonumber \\
&& \ \ \  \ \,   \ \ \  \ \ \ \ \ \  \ \ \  \ \ \  \ \ \   +A_1^2  (256 D^2 \theta_0^8-48 A_3 \theta_0^2 )+192 \theta_0^4  (3 A_2^2-4 A_3 \theta_0^2 ) \big] \Big\}\bigg\}\varepsilon ^2+\cdots. 
\end{eqnarray}
The terms up to the fourth order are too long and therefore we omit them 
for convenience. However, calculating $\mu$ for General Relativity and for 
the covariant $f(T)$ gravity, and finding their difference,  leads to the simple expression
\begin{align}
 \Delta\mu=\mu_{f(T)}-\mu_{GR} \approx
 \frac{64 \alpha \left(2 \theta_0^2+1\right)}{15 \left(\theta_0^2-1\right) \left(\theta_0^2+1\right)^3 M^2} \varepsilon^4 +\mathcal{O}(\varepsilon^5) ~.
\end{align}
Usually one uses the net magnification of flux in the two images to represent 
the characteristics of a gravitational lens 
\cite{Bartelmann:1999yn, Hoekstra:2008db, Bartelmann:2010fz, Cunha:2018acu}. Hence, the total magnification is:
\begin{align}
 \mu_{\mathrm{tot}} = \left|\mu^{+}\right|+\left|\mu^{-}\right| ~,
\end{align}
where $\mu^{+}$ and $\mu^{-}$ are the corresponding magnifications of the two 
images $\theta^{+}$ and $\theta^{-}$. Furthermore, it is more convenient to 
use $\beta$ in order to express total magnification. Hence, we finally arrive at the following expression for the total magnification between the two theories:
\begin{align}
 \Delta \mu_{\mathrm{tot}} = \mu_{\mathrm{tot}}{}_{f(T)} -\mu_{\mathrm{tot}}{}_{GR} \approx -\frac{64 \left( \beta ^4 +6 \beta ^2 +6 \right) \alpha}{15 \beta  \left(\beta ^2+4\right)^{3/2} M^2} \varepsilon^4 +\mathcal{O}(\varepsilon^5) ~.
\end{align}
This is the main result of the present work. It provides the difference in the lensing features between the covariant, consistent, $f(T)$ gravity and General Relativity. Hence, one could use lensing data in order to find the observational signature of $f(T)$ gravity, and extract constraints in the allowed  modification. We mention that in general lensing scales are not  restricted as in the case of Solar System data, and therefore deviations from General Relativity could be observed more easily.

\section{Conclusions}\label{sec:conclusion}

In this work we calculated the deflection angle in the case of covariant 
$f(T)$ gravity, as well as the deviations in the positions and magnifications 
of the lensed images comparing to General Relativity. Firstly, we extracted the 
spherically symmetric solutions for the pure-tetrad as well as the covariant formulation of the theory. Considering spherical solutions, the extension to the latter is crucial, since it ensures that the results do not suffer from 
frame-dependent artifacts that appeared in previous studies of such solutions 
in pure-tetrad formalism. As we indeed saw, the metric behavior for two 
specific tetrad choices corresponding to the same spherically symmetric metric ansatz
is different, which is a not a desirable effect. On the other hand, the 
solution of the covariant case is perfectly self-consistent, independently of 
the tetrad ansatz. Moreover, applying the weak-field limit and perturbative 
approximation we extracted the deviation of the solution comparing to General 
Relativity, which turned out to be at the order of $\frac{1}{r^5}$.

As a next step, we used the obtained solution in order to calculate the 
deflection angle. Expanding the tetrad functions we extracted the difference in 
the light bending comparing to General Relativity, showing that it is of the 
order of $\left(\frac{M}{b}\right)^5$ with $M$ the mass of the  compact body  
and $b$ the impact parameter. Finally, applying the expression for the 
deflection angle in the lensing framework, we calculated the differences   of 
the  positions and magnifications of the lensed images comparing to General 
Relativity.

The effect of $f(T)$ gravity on the lensing features can serve as an observable signature of $f(T)$ gravity in the realistic cases where $f(T)$ is expected to deviate only slightly from General Relativity. In particular, contrary to the Solar System application of regular $f(T)$ gravity, which is impossible to find deviations from General Relativity, the lensing scales in general are not  restricted, and therefore deviations from General Relativity could be observed more easily. It would be interesting to apply these results with galaxy-galaxy weak-lensing datasets (similarly to the analysis of \cite{Chen:2019ftv} for the simple, pure-tetrad version of the theory), as well as in the environments of supermassive black holes (in analogy with \cite{Li:2019lsm} as an application in the Event Horizon Telescope observations) and other lensing datasets, 
in order to extract specific constraints on covariant $f(T)$ gravity. Such a 
full observational confrontation lies beyond the scope of the present work and it is left for a future project.

\acknowledgments
We are grateful to Chao Chen,  Zhaoting Chen, Yiqi Huang, Martin Krssak, Chunlong Li, Jiajun Zhang and Jun Zhang for helpful discussions.
This work is supported in part by the NSFC (Nos. 11653002, 11961131007, 11722327, 11421303), by the CAST Young Elite Scientists Sponsorship (2016QNRC001), by the National Youth Talents Program of China, by the Fundamental Research Funds for Central Universities, by CAS project for young scientists in basic research (YSBR-006), by the CSC Innovation Talent Funds, and by the USTC Fellowship for International Cooperation. 
All numerics were operated on the computer clusters {\it LINDA} \& {\it JUDY} in the particle cosmology group at USTC.

\bibliography{reference}{}
\bibliographystyle{apsrev4-1}

\end{document}